\documentclass[12pt]{iopart}
\usepackage{hyperref}
\usepackage{amssymb}
\usepackage{iopams}

\begin{document}
\title[Cosmological particle creation in Weyl geometry]{Cosmological particle creation in Weyl geometry}
\author{V A Berezin and V I Dokuchaev}
\address{Institute for Nuclear Research of the Russian Academy of Sciences, \\
60th October Anniversary Prospect 7a, 117312 Moscow, Russia}
\eads{\mailto{berezin@inr.ac.ru}, \mailto{dokuchaev@inr.ac.ru}}

\begin{abstract}
We investigated the possibility of the homogeneous and isotropic cosmological solution in Weyl geometry, which differs from the Riemannian geometry by adding the so  called Weyl vector. The Weyl gravity is obtained by constructing the gravitational Lagrangian both to be quadratic in curvatures and conformal invariant. It is found that such solution may exist provided there exists the direct interaction between the Weyl vector and the matter fields. Assuming the matter Lagrangian is that of the perfect fluid, we found how such an interaction can be implemented. Due to the existence of quadratic curvature terms and the direct interaction the perfect fluid particles may be created straight from the vacuum, and we found the expression for the rate of their production which appeared to be conformal invariant. In the case of creating  the universe ``from nothing'' in the vacuum state, we investigated the problem,  whether this vacuum may persist or not. It is shown that the vacuum may persist with respect to producing the non-dust matter (with positive pressure), but cannot resist to producing the dust particles. These particles, being non-interactive, may be considered as the candidates for dark matter.
\end{abstract}

\section{Introduction}

The local conformal invariance seems to be a good candidate to become  the fundamental symmetry of the Nature. It looks very plausible if one suppose that the universe was created ``from nothing''\cite{Vilenkin}.  

Nowadays this idea is advocated, among others, by Roger Penrose\cite{Penrose} and Gerard 't Hooft\cite{Hooft}. But the first claim was made by Hermann Weyl\cite{Weyl} a little bit more than one hundred years ago. He tried to construct the unified theory of gravitational and electromagnetic interactions in a pure geometrical way. In order to achieve such a goal, he invented the new geometry, which we call now the Weyl geometry. It differs from the Riemannian geometry by adding to the metric tensor some vector (1-forms).  Hermann Weyl discovered that in order the whole theory to become invariant under the local conformal transformation of the metric tensor, this new (Weyl) vector has to undergo the specific gauge transformation, that is why he decided to consider it as the electromagnetic vector. It is well known that Albert Einstein criticized such a theory (second clock problem and all that) and it was abandoned.  Nevertheless, if one considers the Weyl vector as just the part of the geometry (with no link to the electromagnetism), then the Weyl gravity remains the beautiful example of the non-Riemannian conformal invariant theory.

Our interest in cosmological solution in the Weyl gravity arises from the fact that requirement of the local conformal invariance in the Riemannian geometry leads to the conclusion that all the homogeneous and isotopic space-times are just the vacuum solutions. The aim of present paper is to investigate this problem in the framework of Weyl geometry.

\section{Introduction to Weyl geometry}

The differential geometry is completely determined by the metric tensor $g_{\mu\nu}$ and connections
$\Gamma^\lambda_{\mu\nu}(x)$, the former provides us with the interval between the nearby points,
\begin{equation} \label{ds2}
	ds^2=g_{\mu\nu}dx^\mu dx^\nu,
\end{equation}
while the latter serves for defining the parallel transfer of vectors and tensors and their covariant derivatives
\begin{equation} \label{nabla}
\nabla_{\lambda} l^\mu= l^\mu_{\;,\lambda} +\Gamma_{\lambda\nu}^\mu l^\nu\ldots,
\end{equation}
``comma'' denotes a partial derivative.

The curvature tensor $R^{\mu}_{\phantom{\mu}\nu\lambda\sigma}$ is constructed solely of $\Gamma_{\mu\nu}^\lambda$, namely
\begin{equation} \label{curv}
	R^{\mu}_{\phantom{\mu}\nu\lambda\sigma}=\frac{\partial \Gamma^\mu_{\nu\sigma}}{\partial x^\lambda}-\frac{\partial \Gamma^\mu_{\nu\lambda}}{\partial x^\sigma}+\Gamma^\mu_{\varkappa\lambda}\Gamma^\varkappa_{\nu\sigma}-\Gamma^\mu_{\varkappa\sigma}\Gamma^\varkappa_{\nu\lambda},
\end{equation}
as well as its convolution, Ricci tensor $R_{\mu\nu}$,
\begin{equation} \label{Ricci}
	R_{\mu\nu}=R^{\lambda}_{\phantom{\mu}\mu\lambda\nu}.
\end{equation}
The curvature scalar is $R=g^{\mu\nu}R_{\mu\nu}$.

It appears possible to calculate them, if one knows three tensors, the metric tensor $g_{\mu\nu}$, the torsion
$S^\lambda_{\phantom{\mu}\mu\nu}$ and the nonmetricity $Q_{\lambda\mu\nu}$, where
\begin{equation} \label{S}
S^\lambda_{\phantom{\mu}\mu\nu}=\Gamma^\lambda_{\mu\nu} -\Gamma^\lambda_{\mu\nu},
\end{equation}
\begin{equation} \label{Q}
Q_{\lambda\mu\nu}= \nabla_\lambda g_{\mu\nu},
\end{equation}
namely
\begin{equation} \label{Gamma}
\Gamma^\lambda_{\mu\nu}=C^\lambda_{\phantom{\mu}\mu\nu} + K^\lambda_{\phantom{\mu}\mu\nu} +L^\lambda_{\phantom{\mu}\mu\nu},
\end{equation}
where $C^\lambda_{\phantom{\mu}\mu\nu}$ are Christoffel symbols,
\begin{equation} \label{C}
C^\lambda_{\phantom{\mu}\mu\nu}=\frac{1}{2}g^{\lambda\kappa} (g_{\kappa\mu,\nu}+g_{\kappa\nu,\mu}-g_{\mu\nu,\kappa})
\end{equation}
and
\begin{equation} \label{K}
K^\lambda_{\phantom{\mu}\mu\nu}=\frac{1}{2} (S^\lambda_{\phantom{\mu}\mu\nu}-S^{\phantom{\mu}\lambda}_{\mu\phantom{\mu}\nu} -S^{\phantom{\mu}\lambda}_{\nu\phantom{\mu}\mu}),
\end{equation}
\begin{equation} \label{L}
L^\lambda_{\phantom{\mu}\mu\nu}=\frac{1}{2} (Q^\lambda_{\phantom{\mu}\mu\nu}-Q^{\phantom{\mu}\lambda}_{\mu\phantom{\mu}\nu} -Q^{\phantom{\mu}\lambda}_{\nu\phantom{\mu}\mu}).
\end{equation}
Note, that, using the Christoffel symbols as the connections, we can construct another covariant derivative, the metric one, different (in general) from $\nabla_\mu$, which will be denoted by a semicolon ``$;$''. 

The lowering and raising of indices are performed by the metric tensor, $g_{\mu\nu}$, and its inverse $g^{\mu\nu}$ ($g^{\mu\nu}g_{\nu\lambda}=\delta_\lambda^\mu$), correspondingly.

With such a classification scheme the Riemannian geometry appears the simplest one. Indeed, in this case $S^\lambda_{\phantom{\mu}\mu\nu}=0$, $Q_{\lambda\mu\nu}=0$, and the metric tensor describes everything. Moreover, the curvature tensor $R_{\mu\nu\lambda\sigma}$ possesses the additional algebraic symmetries (by definition, it is skew-symmetric in the second pair of indices)
\begin{equation} \label{R2}
R_{\mu\nu\lambda\sigma}=R_{\lambda\sigma\mu\nu} = -R_{\nu\mu\lambda\sigma}=-R_{\mu\nu\sigma\lambda},
\end{equation}
\begin{equation} \label{R3}
R^\mu_{\phantom{\mu}\nu\lambda\sigma} +R^\mu_{\phantom{\mu}\sigma\nu\lambda} +R^\mu_{\phantom{\mu}\lambda\sigma\nu} =0,
\end{equation}
and obeys the Bianchi identities
\begin{equation} \label{Bianchi}
R^\mu_{\phantom{\mu}\nu\lambda\sigma;\kappa}  +R^\mu_{\phantom{\mu}\nu\kappa\lambda;\sigma} + R^\mu_{\phantom{\mu}\nu\sigma\kappa;\lambda}=0.
\end{equation}
Besides, Ricci tensor is symmetrical,
\begin{equation} \label{Ricci2}
R_{\mu\nu}=R_{\nu\mu}.
\end{equation}

The Weyl geometry starts to fill the next level. The torsion tensor is still zero, $S^\lambda_{\phantom{\mu}\mu\nu}=0$ ($\Gamma^\lambda_{\mu\nu}=\Gamma^\lambda_{\nu\mu}$), but the nonmetricity is not,
\begin{equation} \label{Q2}
Q_{\lambda\mu\nu}= \nabla_\lambda g_{\mu\nu} =A_\lambda g_{\mu\nu},
\end{equation}
the 1-form $A_\lambda$ is called the Weyl vector. Accordingly, the connections are
\begin{equation} \label{Gamma2}
\Gamma^\lambda_{\mu\nu}=C^\lambda_{\mu\nu}+W^\lambda_{\mu\nu},
\end{equation}
\begin{equation} \label{W}
W^\lambda_{\mu\nu}=-\frac{1}{2}(A_\mu \delta^\lambda_\nu+ A_\nu \delta^\lambda_\mu -A^\lambda g_{\mu\nu}),
\end{equation}
where $\delta^\lambda_\nu$ is the Kronecker symbol. 

The curvature tensor $R_{\mu\nu\lambda\sigma}$, is, of course, skew-symmetric in the second pair of indices, but some other symmetries are lost. Instead, 
\begin{equation} \label{R3}
R_{\mu\nu\lambda\sigma}+R_{\nu\mu\lambda\sigma}=2F_{\lambda\sigma}g_{\mu\nu},
\end{equation}
\begin{eqnarray} \label{R3}
R_{\mu\nu\lambda\sigma}-R_{\lambda\sigma\mu\nu}&=& \frac{1}{2}\left(F_{\mu\sigma}g_{\nu\lambda} -F_{\mu\lambda}g_{\nu\sigma} +F_{\nu\lambda}g_{\mu\sigma} \right. \nonumber \\ &&-\left.F_{\nu\sigma}g_{\mu\lambda}+F_{\mu\nu}g_{\lambda\sigma} -F_{\lambda\sigma}g_{\mu\nu}\right).
\end{eqnarray}
In spite of this, the cyclic sum remains zero
\begin{equation} \label{cyclic}
R_{\mu\nu\lambda\sigma} +R_{\mu\sigma\nu\lambda} +R_{\mu\lambda\sigma\nu} =0.
\end{equation}
The Ricci tensor is no more symmetric,
\begin{equation} \label{Ricci3}
	R_{\nu\sigma}-R_{\sigma\nu}=F_{\nu\sigma}.
\end{equation}
Here $F_{\mu\nu}$ is the strength tensor,
\begin{equation} \label{Stress}
F_{\mu\nu}=\nabla_\mu A_\nu - \nabla_\nu A_\mu =A_{\nu;\mu} - A_{\mu;\nu} =A_{\nu,\mu} - A_{\mu,\nu}.
\end{equation}

In 1919 Hermann Weyl proposed the unification of the electromagnetic and gravitational interactions based on such a geometry.

\section{Local conformal transformation and Weyl gravity}

Hermann Weyl claimed that the gravitation should be invariant under the local conformal transformation
\begin{equation} \label{local}
ds^2=\Omega^2(x)d\hat s^2=\Omega^2(x)\hat g_{\mu\nu}dx^\mu dx^\nu,
\end{equation}
like the classical electrodynamics. Here $\Omega(x)$ is the conformal factor, and we denote by ``hats'' the conformally  transformed quantities. Note, that the conformal transformation does not change the coordinates. Christoffel symbols are transformed, evidently, in the following way
\begin{eqnarray} \label{hatC}
C^\lambda_{\mu\nu}&=&\frac{1}{2\Omega^2}\hat g^{\lambda\sigma}\left((\Omega^2\hat g_{\sigma\mu})_{,\nu} +(\Omega^2\hat g_{\sigma\nu})_{,\mu} -(\Omega^2\hat g_{\mu\nu})_{,\sigma}\right)
\nonumber \\ 
&=&\hat C^\lambda_{\mu\nu} +\left(\frac{\Omega_{,\mu}}{\Omega}\delta^\lambda_\nu +\frac{\Omega_{,\nu}}{\Omega}\delta^\lambda_\mu -\frac{\Omega_{,\kappa}}{\Omega}\hat g^{\lambda\kappa}\hat g_{\mu\nu}\right).
\end{eqnarray}
Following Weyl's idea, let us postulate that 
\begin{equation} \label{hatA}
	\Gamma^\lambda_{\mu\nu}=\hat\Gamma^\lambda_{\nu\mu}.
\end{equation}
Then, since $\Gamma^\lambda_{\mu\nu}=C^\lambda_{\mu\nu}+W^\lambda_{\mu\nu}$,
\begin{eqnarray} \label{26}
&&C^\lambda_{\mu\nu}-\frac{1}{2}(A_\mu \delta^\lambda_\nu +A_\nu \delta^\lambda_\mu -A^\lambda g_{\mu\nu})
\nonumber \\ 
&=&\hat C^\lambda_{\mu\nu} -\frac{1}{2}(\hat A_\mu \delta^\lambda_\nu +\hat A_\nu \delta^\lambda_\mu -\hat g^{\lambda\sigma}\hat A_\sigma \hat g_{\mu\nu}).
\end{eqnarray}
Substituting the Eq. (\ref{hatC}) and making the convolution, say over $\lambda=\nu$, one gets readily
\begin{equation} \label{hatA}
A_\mu=\hat A_\mu+2\frac{\Omega_{,\mu}}{\Omega}.
\end{equation}
It would become the gauge field, like in the classical electrodynamics. 

Since both curvature and Ricci tensors are constructed solely from $\Gamma^\lambda_{\mu\nu}$, it is evident that
\begin{equation} \label{Rhat2}
R^\mu_{\phantom{1}\nu\lambda\sigma}=\hat R^\mu_{\phantom{1}\nu\lambda\sigma},
\end{equation}
\begin{equation} \label{Rhat3}
R_{\mu\nu}=\hat R_{\mu\nu}.
\end{equation}
Trying to incorporate the electrodynamics into the geometry, like the gravitation, and to construct the gravitational Lagrangian, like the electrodynamics, H.Weyl postulated the following conformal invariant action integral 
\begin{equation} \label{SW}
S_{\rm W} =\int\!{\cal L_{\rm W}}\sqrt{-g}\,d^4x,
\end{equation}
\begin{equation} \label{LW}
{\cal L_{\rm W}}=\alpha_1  R_{\mu\nu\lambda\sigma}R^{\mu\nu\lambda\sigma}
+\alpha_2R_{\mu\nu}R^{\mu\nu}+\alpha_3R^2 +\alpha_4F_{\mu\nu}F^{\mu\nu}.
\end{equation}
Note that the curvature scalar $R$ is not conformal invariant, the conformal invariant combination is $R^2\sqrt{-g}$. We will call this ``the Weyl gravity''. 

The total action integral consists of the gravitational part, $S_{\rm W}$, and the action for the matter fields, $S_{\rm m}$,
\begin{equation} \label{total}
S_{\rm tot}=S_{\rm W}+S_{\rm m}, \quad S_{\rm m}= \int\!{\cal L_{\rm m}}\sqrt{-g}\,d^4x.
\end{equation}
Indeed, to get the conformal invariant equations of motion it is sufficient to have the conformal invariant variations, $\delta S_m$ of the matter action integral. Thus, the requirement for total action to be conformal invariant seems too strong. It is important to note, that, while the Weyl action is conformal invariant, the matter one, $S_{\rm m}$, is not necessarily has such a property, but, its variation, $\delta S_{\rm m}$, is required to be conformal invariant. Indeed, to get the conformal invariant equations of motion it is sufficient to have the conformal invariant variations, $\delta S_{\rm m}$ of the matter action integral. Thus, the requirement for total action to be conformal invariant seems too strong.

By definition,
\begin{eqnarray} \label{R3}
\delta S_{\rm m}\stackrel{\mathrm{def}}{=}&& -\frac{1}{2}\!\int\! T^{\mu\nu}(\delta g_{\mu\nu})\sqrt{-g}\,d^4x
-\!\int\!\! G^\mu(\delta A_\mu)\sqrt{-g}\,d^4x
 \nonumber \\ 
&&+\int\!\frac{\delta \cal L_{\rm  W}}{\delta\Psi} (\delta \Psi)\sqrt{-g}\,d^4x,
\end{eqnarray}
where $G^\mu$ is some vector that can be called ``the Weyl current'', and $\Psi$ is the collective dynamical variable, describing the matter field, the matter motion being determined by the Lagrange equation $\delta S/\delta\Psi=0$.

Since
\begin{equation} \label{deltag}
\delta g_{\mu\nu}=2\Omega\hat g_{\mu\nu}(\delta\Omega)=2g_{\mu\nu}\frac{\delta\Omega}{\Omega},
\end{equation}
\begin{equation} \label{deltaOmega}
\delta A_\mu=2\delta\left(\!\frac{\Omega_{,\mu}}{\Omega}\!\right) = 2\delta(\log\Omega)_{,\mu} =2(\delta(\log\Omega))_{,\mu} =2\left(\!\frac{\delta\Omega}{\Omega}\!\right)_{\!\!,\mu},
\end{equation}
then,
\begin{equation} \label{deltaOmega}
\delta S= 0 = -\!\int\!T^{\mu\nu}g_{\mu\nu}\left(\!\frac{\delta\Omega}{\Omega}\!\right)\!\sqrt{-g}\,d^4x - 2\int\!G^\mu \left(\!\frac{\delta\Omega}{\Omega}\!\right)_{\!\!,\mu} \!\!\sqrt{-g}\,d^4x,
\end{equation}
and one obtains, after removing  the full derivative,   
\begin{equation} \label{2G}
2(G^\mu)_{\;;\mu}={\rm Trace}T^{\mu\nu}.
\end{equation}
This can be called ``the self-consistency condition''. Note, that it contains not the Weyl covariant derivative, but the metric one. The self-consistency condition must be added to the field equations, which we  will write down later for the homogeneous and isotropic cosmological space-times.

\section{Perfect fluid}

In what follows we will consider the perfect fluid as the matter field  and choose for its action integral the following one\cite{Ray},
\begin{eqnarray} \label{R3}
S_{\rm m}&=& -\!\int\!\varepsilon(X,n)\sqrt{-g}\,d^4x + \int\!\lambda_0(u_\mu u^\mu-1)\sqrt{-g}\,d^4x
\nonumber \\ 
&&+\int\!\lambda_1(n u^\mu)_{;\mu}\sqrt{-g}\,d^4x + \int\!\lambda_2 X_{,\mu}u^\mu\sqrt{-g}\,d^4x.
\end{eqnarray}
The dynamical variables are the particle number density $n(x)$, the four-velocity $u^\mu(x)$ and the auxiliary variable $X(x)$.

The corresponding equations of motion are  the following
\begin{eqnarray} \label{eqmot}
\delta n:&& -\frac{\partial\varepsilon}{\partial n}-\lambda_{1,\sigma} u^\sigma=0,
\\ 
\delta u^\mu:&& -2\lambda_0 u_\mu -\lambda_{1,\mu}n +\lambda_2 X_{,\mu}=0,
\\ 
\delta X:&& -\frac{\partial\varepsilon}{\partial X}-(\lambda_2u^\sigma)_{;\sigma} =0.
\end{eqnarray}
The variation of the Lagrange multipliers $\lambda_0$, $\lambda_1$ and $\lambda_2$ i,pose the constraints, the four velocity normalization $u^\mu u_\mu=1$, the particle number conservation $(nu^\mu)_{;\mu}=0$ and the enumeration of trajectories $X_{,\mu}u^\mu=0$, respectively. 

Contracting the second equation of motion with $u^\mu$ and using the constraints, one gets
\begin{equation} \label{lambda0a}
2\lambda_0=n\lambda_{1,\sigma}u^\sigma=-n\frac{\partial\varepsilon}{\partial n},
\end{equation}
the last equality follows from the first (scalar) equation of motion. Introducing the hydrodynamical pressure $p$ by the usual relation
\begin{equation} \label{p}
p=n\frac{\partial\varepsilon}{\partial n} -\varepsilon,
\end{equation}
one obtains, eventually,
\begin{equation} \label{pe}
2\lambda_0=-(\varepsilon+p).
\end{equation}

The variation of the metric tensor gives us, by definition, the energy-momentum tensor $T^{\mu\nu}$.
\begin{eqnarray} \label{eqmot}
\delta S_{\rm m}&\stackrel{\mathrm{def}}{=}& 
-\frac{1}{2}\!\int\! T^{\mu\nu}(\delta g_{\mu\nu})\sqrt{-g}\,d^4x 
\nonumber \\ 
&=&-\frac{1}{2}\!\int\!\varepsilon g^{\mu\nu}(\delta g_{\mu\nu})\sqrt{-g}\,d^4x +\!\int\!\lambda_0 u^\mu u^\nu(\delta g_{\mu\nu})\sqrt{-g}\,d^4x,
\nonumber \\ 
&&-\frac{1}{2}\!\int\!n\lambda_{1,\sigma} u^\sigma g^{\mu\nu}(\delta g_{\mu\nu})\sqrt{-g}\,d^4x,
\end{eqnarray}
where we already made use of the constraints. Hence,
\begin{equation} \label{T0}
	T^{\mu\nu}=\varepsilon g^{\mu\nu} -2\lambda_0 u^\mu u^\nu +n\lambda_{1,\sigma}u^\sigma g^{\mu\nu}.
\end{equation}
Substituting the expressions for $\lambda_0$ and $\lambda_{1,\sigma}u^\sigma$, one arrives at the famous expression
\begin{equation} \label{T}
T^{\mu\nu}=(\varepsilon+p)u^\mu u^\nu -p g^{\mu\nu}.
\end{equation}

Our aim is to study possible modifications of the perfect fluid action integral due to the transition from the Riemannian to the Weyl geometry. To do this, let us consider the behavior of single particle in the given gravitational field. It is very well known that in the case of Riemannian geometry the only possible choice is
\begin{equation} \label{part}
S_{\rm part} = -m\!\int\!ds=-m\!\int\!\sqrt{g^{\mu\nu}(x)\frac{d x^\mu}{d\tau}\frac{d x^\nu}{d\tau}}d\tau,
\end{equation}
where $m$ is the rest mass of the particle, $\tau$ is its proper time, and the dynamical variable is the trajectory $x^\mu(\tau)$. Then, the least action principle $\delta S_{\rm part}=0$ leads to the geodesic motion, $u_{\mu;\nu}u^\nu=0$ ($u^\mu=dx^\mu/d\tau$).

But in the case of the Weyl geometry there exists yet another invariant, $B$,
\begin{equation} \label{B}
B=A_\mu u^\mu, \quad u^\mu=\frac{dx^\mu}{d\tau}.
\end{equation}
The possible structure of the action integral becomes more complicated
\begin{equation} \label{Spart2} 
	S_{\rm part} =\!\int\!\!f_1(B)ds \! +\!\int\!\!f_2(B)d\tau = \!\int\!\!\left\{\right.f_1(B)\sqrt{g_{\mu\nu}u^\mu u^\nu} \! + f_2(B)\left.\right\}d\tau, 
\end{equation}
with some arbitrary functions $f_1(B)$ and $f_2(B)$. The corresponding equations of motion are
\begin{eqnarray} \label{R3}
f_1(B)u_{\lambda;\mu}u^\mu&=&\left( (f_1^{''}(B)+f_2^{''}(B))A_\lambda- f_1^{'}(B)u_\lambda\right)B_{,\mu}u^\mu 
\nonumber \\ 
&&+ (f_1^{'}(B)+f_2^{'}(B))F_{\lambda\mu}u^\mu,
\end{eqnarray}
\begin{equation} \label{Spart2}
 F_{\lambda\mu}=A_{\mu,\lambda}-A_{\lambda,\mu}.
\end{equation}
Since $F_{\lambda\mu}u^\lambda u^\mu\equiv0$ and 
$u_{\lambda;\mu}u^\lambda\equiv0$, the above equations are self-consistent if either
\begin{equation} \label{f1f2}
(f_1^{''}+f_1^{''})B-f_1^{'}=0
\end{equation}
or
\begin{equation} \label{Bu0}
	B_{,\mu}u^\mu=0.
\end{equation}
	
How to insert the interaction with the Weyl vector $A_\mu$ into the perfect fluid Lagrangian?  Evidently, the new invariant $B=A_\mu u^\mu$ is tightly linked to the particle number density $n$. So, the simplest way to make the replacement
\begin{equation} \label{nB}
n \;\; \rightarrow \;\; \varphi(B)n, \quad \varepsilon=\varepsilon(X,\varphi(B)n).
\end{equation}
Here $\varphi(B)$ is some arbitrary function of this new invariant with $\varphi(0)\neq0$. This causes the contributions, $G^\mu[\rm part]$, to the Weyl current and to the energy-momentum tensor, $T^{\mu\nu}[\rm part]$, namely 
\begin{equation} \label{Gpart}
G^\mu[\rm part]=\frac{\varphi'(B)}{\varphi(B)}(\varepsilon+p)u^\mu,
\end{equation}
\begin{equation} \label{Tpart}
T^{\mu\nu}[\rm part]=(\varepsilon+p) \left(1-B\frac{\varphi'(B)}{\varphi(B)}\right)u^\mu u^\nu -pg^{\mu\nu}.
\end{equation}

How about the particle number conservation law, $(nu^\mu)_{;\mu}=0$?

\section{Particle production rate}

The investigation of process of particle creation from vacuum fluctuations of scalar fields \cite{ZeldStarob,ParkerFulling,GribMamMostep}  showed the main role played by the so called conformal anomalies. The local part of them can be described by inclusion into the gravitational action integrals some combination of the terms quadratic in curvatures (in the one-loop approximation of the quantum field theory). Since the Lagrangian of Weyl gravity consists just of such quadratic terms, the particle creation must take place. Thus, we have to have
\begin{equation} \label{Phiinv}
(nu^\mu)_{;\mu}=\Phi(\rm inv),
\end{equation}
Note, that the covariant derivative here is the metric covariant derivative, associated with the Christoffel symbols. Only in this case the volume integral of the left-hand-side is completely converted into the surface integral and equals the particle number flow across the boundary. ``The law of creation'' $\Phi$ depends on some invariants. The simplest way to take this into account in the perfect fluid Lagrangian is to modify the corresponding constraint\cite{Berezin}.

Thus we arrive at the following matter action integral
\begin{eqnarray} \label{R3}
S_{\rm m}&=& -\int\!\varepsilon(X,\varphi(B)n)\sqrt{-g}\,d^4x + \int\!\lambda_0(u_\mu u^\mu-1)\sqrt{-g}\,d^4x
\nonumber \\ 
&+&\int\!\!\lambda_1\left((nu^\mu)_{;\mu}-\Phi(\rm inv)\right)\sqrt{-g}\,d^4x +\! \int\!\lambda_2 X_{,\mu}u^\mu\sqrt{-g}\,d^4x.
\end{eqnarray}

Since the Weyl gravity is conformal invariant it seems natural to check the behavior of the creation function $\Phi$ under such a transformation. One has
\begin{equation} \label{nhat}
n=\frac{\hat n}{\Omega^3},  \quad u^\mu=\frac{\hat u^\mu}{\Omega},  \quad \sqrt{-g}=\Omega^4 \sqrt{-\hat g},
\end{equation}
hence
\begin{eqnarray} \label{numu2}
(n u^\mu)_{;\mu}&=&\frac{1}{\sqrt{-g}}(n u^\mu\sqrt{-g})_{,\mu}= \frac{1}{\sqrt{-g}}\left(\frac{\hat n}{\Omega^3} \frac{\hat u^\mu}{\Omega}\Omega^4\sqrt{-\hat g}\right)_{,\mu}=
\nonumber \\ 
&=& \frac{1}{\sqrt{-g}}(\hat n\hat  u^\mu\sqrt{-\hat g})_{,\mu},
\end{eqnarray}
and we obtain, that $(n u^\mu)_{;\mu}\sqrt{-g}$ is conformal invariant. It is not at all surprising because the number of particles can be just counted.

Let us assume that there are no classical fields, and the particle are created solely by the vacuum fluctuations (i.\,e., by geometry). 

In the case of the Weyl geometry we are ready to write down the result for the creation law,
\begin{equation} \label{PhiWeyl}
\Phi=\alpha_1'R_{\mu\nu\lambda\sigma}R^{\mu\nu\lambda\sigma}
+\alpha_2'R_{\mu\nu}R^{\mu\nu}
+\alpha_3'R^2+ \alpha_4'F_{\mu\nu}F^{\mu\nu}
\end{equation}
(for the reasons described above, we are interested only in quadratic terms). This equation is, surely, not unique. Our choice was dictated by the assumption that classical fields are absent (only the geometry), and the restriction to the quadratic terms in curvatures was justified by the reference to the Weyl geometry. Surely, $\Phi$ will contribute both to the Weyl current ($G^\mu[\rm cr]$) and to the energy-momentum tensor ($T^{\mu\nu}[\rm cr]$).

It is widely known that in the Riemannian geometry the only conformal invariant combination which is quadratic in curvatures is the square of the Weyl tensor, so
\begin{equation} \label{RiemannianC2}
	(nu^\mu)_{;\mu}=\eta C^2,
\end{equation}
where  $\eta=const$, and $C^2$ is the square of the Weyl tensor $C_{\mu\nu\lambda\sigma}$.
Exactly the same result were obtained by Ya. B. Zel'dovich and A. A. Starobinski in 1977 \cite{ZeldStarob} while they studied the particle creation by the vacuum fluctuations of the massless scalar field on the background metric of the homogeneous and slightly anisotropic cosmological space-time obeying the Einstein equations. Now it becomes fundamental for any Reemannian geometry, irrespective of the form of gravitational Lagrangian.

\section{Cosmology}

By cosmology we understand the homogeneous and isotropic space-times described by the Robertson-Walker metric,
\begin{equation} \label{RW}
ds^2=dt^2-a^2(t)dl^2,
\end{equation}
\begin{equation} \label{RW}
dl^2=\gamma_{ij}dx^idx^j=\frac{dr^2}{1-kr^2}+r^2(d\theta^2+\sin^2\theta d\varphi^2), \quad (k=0,\pm1),
\end{equation}
with the scale factor $a(t)$.

Due to the high level of the symmetry one has, in the Robertson-Walker metric,
\begin{equation} \label{A4}
A_\mu=(A_0(t),0,0,0), 
\end{equation}
from where it follows, that
\begin{equation} \label{F0}
F_{\mu\nu}\equiv0.
\end{equation}
Also
\begin{equation} \label{Tmunu}
T_\mu^\nu=(T_0^0,T_1^1=T_2^2=T_3^3)= T_\mu^\nu(t).
\end{equation}
Clearly $T_1^1=(1/3)(T-T_0^0)$, $T={\rm Trace}T^{\mu\nu}$ and we need to know only $T_0^0$ and $T$.

Since $A_0(t)=\hat  A(t) +2(\dot\Omega/\Omega)$, it is always possible to find a gauge with $\hat A(t)=0$, which we will call ``the special gauge'', and the corresponding solutions may be called ``the basic solutions''. Side note: $B=A_\mu u^\nu=0$, and all the functions of $B$ are converted into the set of some constants.

We are not allowed to put $A_\mu=0$ prior to the variation process because $\delta A_\mu\neq0$. Hence, one should extract first the variation $\delta S/\delta A_\mu$ and evaluate $G^\mu[\rm cr]$,
\begin{equation} \label{TntGmucr}
\int G^\mu[\rm cr](\delta A_\mu)\sqrt{-g}\,d^4x =\int\lambda_1(x)\frac{\delta\Phi(\rm inv)}{\delta A_\mu}(\delta A_\mu)\sqrt{-g}\,d^4x.
\end{equation}
The left-hand-side of the corresponding gravitational equation  will then be obtained simply by putting $\lambda_1=1$ (and without ``primes''). The detailed calculations will be presented in the Appendices A and B, here we give the final result, 
\begin{eqnarray} \label{Gmucr}
G^\mu[\rm cr]&=&-2(2\alpha_1'+\alpha_2')\lambda_{1;\kappa}R^{\mu\kappa}
-(\alpha_2'+6\alpha_3')\lambda_1^{;\mu}R
\nonumber \\ 
&&-2(\alpha_1' +\alpha_2'+3\alpha_3')\lambda_1R^{;\mu},
\end{eqnarray}
where we already took into account that for any homogeneous and isotropic space-time  the Weyl tensor $C_{\mu\nu\lambda\sigma}\equiv0$. In cosmology 
$R^{00}=R_0^0(t)$, $R^{i0}=0$, $R=R(t)$, $R^{ij}=R_1^1g^{ij}$, $R_1^1=(1/3)(R-R_0^0)$,
and one has
\begin{equation} \label{Gi}
G^i[\rm cr]=0,
\end{equation}
\begin{eqnarray} \label{G0cr}
G^0[\rm cr] &=&-2(2\alpha_1'+\alpha_2')\dot\lambda_1R^0_0
-(\alpha_2'+6\alpha_3')\dot\lambda_1R
\nonumber \\ 
&&-2(\alpha_1'+\alpha_2'+3\alpha_3')\lambda_1\dot R.
\end{eqnarray}

Now we can safely jump to our special gauge $A=0$, where we will be dealing, essentially, with the Riemannian geometry. The gravitational Lagrangian, ${\cal{L}_{\rm  W}}$, becomes
\begin{eqnarray} \label{LW}
{\cal{L}_{\rm  W}} &=&\alpha_1R_{\mu\nu\lambda\sigma}R^{\mu\nu\lambda\sigma} +\alpha_2R_{\mu\nu}R^{\mu\nu} +\alpha_3R^2 +\alpha_4F_{\mu\nu}F^{\mu\nu}
\nonumber \\ 
&=& \alpha C^2+\beta GB+\gamma R^2 +\alpha_4F_{\mu\nu}F^{\mu\nu},
\end{eqnarray} 
where
$C^2=C_{\mu\nu\lambda\sigma}C^{\mu\nu\lambda\sigma}$, $C_{\mu\nu\lambda\sigma}$ \ --- \ Weyl tensor, which is the completely traceless part of  the curvature tensor
\begin{eqnarray} \label{C4}
C_{\mu\nu\lambda\sigma} &=&R_{\mu\nu\lambda\sigma}
-\frac{1}{2}R_{\mu\lambda}g_{\nu\sigma} +\frac{1}{2}R_{\mu\sigma}g_{\nu\lambda}
+\frac{1}{2}R_{\nu\lambda}R_{\mu\sigma}
\nonumber \\ 
&& -\frac{1}{2}R_{\nu\sigma}R_{\mu\lambda}
+\frac{1}{6}R(g_{\mu\lambda}g_{\nu\sigma} -g_{\mu\sigma}g_{\nu\lambda}),
\end{eqnarray} 
GS is the Gauss-Bonnet term,
\begin{equation} \label{GB}
GB=R_{\mu\nu\lambda\sigma}R^{\mu\nu\lambda\sigma} -4R_{\mu\nu}R^{\mu\nu} +R^2,
\end{equation}
and 
\begin{eqnarray} \label{C4}
\alpha_1&=&\alpha+\beta \\ 
\alpha_2&=&-2\alpha-4\beta \\ 
\alpha_3&=&\frac{1}{3}\alpha+\beta+\gamma.
\end{eqnarray} 
One can put $C^2=0$ and $F_{\mu\nu}F^{\mu\nu}=0$ straight in the Lagrangian, because $C_{\mu\nu\lambda\sigma}(\delta C^{\mu\nu\lambda\sigma})=0$ and $F^{\mu\nu}(\delta F_{\mu\nu})=0$. Moreover, in 4-dimensional space-times the Gauss-Bonnet term is a full derivative, so it has no effect on the field equations. Thus we are left with only one term, $\gamma R^2$. But in the creation function $\Phi(\rm inv)$ we can not neglect the Gauss-Bonnet term, because it enters the matter Lagrangian with the multiplier $\lambda_1(x)$.

Let us rewrite $G^0[\rm cr]$ using new set of coupling constants, 
\begin{equation} \label{GB}
G^0[\rm cr]=4\beta\dot\lambda_1R^0_0 -2(\beta+3\gamma)\dot\lambda_1R -6\gamma\lambda_1\dot R.
\end{equation}
We see that $\alpha'$ does not appear, as it should be, and dependence on $\beta'$ disappears for $\lambda_1=const$, as it should be.

Finally, we write down the result for $T[\rm cr]$ and $T^0_0[\rm cr]$ (for detailed calculations see Appendix B),
\begin{eqnarray} \label{Tcr}
T[\rm cr]&=&\ddot\lambda_1(8\beta'R^0_0 -4\beta'R -12\gamma'R) \nonumber \\ 
&&-4\dot\lambda_1\left(\beta'\frac{\dot a}{a}(R+2R^0_0)+6\gamma'\dot R +9\gamma'\frac{\dot a}{a}R\right) \nonumber  \\ 
&&-12\lambda_1\gamma'(\ddot R+3\frac{\dot a}{a}\dot R).
\end{eqnarray} 
\begin{eqnarray} \label{T00cr}
T^0_0[\rm cr]&=&8\gamma'\dot\lambda_1\frac{\dot a}{a}R_0^0 -4(\beta'+3\gamma')\dot\lambda_1\frac{\dot a}{a}R \nonumber \\ 
&&-\gamma'\lambda_1\left(12\frac{\dot a}{a}\dot R +R(4R^0_0-R)\right).
\end{eqnarray} 

We are ready now to present the complete set of equations for the cosmological space-times filled with the perfect fluid.

Vector:
\begin{equation} \label{Vector}
-6\gamma\dot R=G^0.
\end{equation}
Tensor:
\begin{equation} \label{Vector00}
-\gamma\left(12\frac{\dot a}{a}\dot R+ R(4R^0_0-R)\!\right)=T^0_0,
\end{equation}
\begin{equation} \label{VectorT}
-12\gamma\left(\ddot R+ 3\frac{\dot a}{a}R\!\right)=T.
\end{equation}
Self-consistency condition:
\begin{equation} \label{Selfc}
2\frac{(G^0a^3)^{\dot{}}}{a^3}=T^0_0+3T^1_1=T.
\end{equation}
It is quite clear that the self-consistency  condition is just the consequence of the vector and trace equations. 

Ricci tensor and scalar curvature are
\begin{equation} \label{Ricci00}
R^0_0=-3\frac{\ddot a}{a},
\end{equation}
\begin{equation} \label{Ricci4}
R=-6\left(\frac{\ddot a}{a}+\frac{\dot a^2+k}{a^2}\right), \quad k=0,\pm1.
\end{equation}
Of all the perfect fluid equations of motion only two are survived,
\begin{equation} \label{survived1}
\dot\lambda_1=-\frac{\varepsilon+p}{n},
\end{equation}
\begin{equation} \label{survived2}
\frac{(na^3)^{\dot{}}}{a^3}=\Phi(\rm inv),
\end{equation}
where
\begin{equation} \label{PhiInv}
\Phi(\rm inv)=-\frac{4}{3}\beta'R^0_0(2R^0_0-R)+\gamma'R^2.
\end{equation}
Remember that
\begin{equation} \label{G0rem}
G^0=G^0[\rm part]+G^0[\rm cr], \quad G^0[\rm part] =\frac{\dot\varphi(0)}{\varphi(0)}(\varepsilon+p),
\end{equation}
\begin{equation} \label{T0rem}
T^0_0=T^0_0[\rm part]+T^0_0[\rm cr], \quad T^0_0[\rm part]=\varepsilon(\varphi(0)n),
\end{equation}
\begin{equation} \label{Trem}
	T=T[\rm part]+T[\rm cr], \quad T[\rm part]=\varepsilon-3p, \quad (p =n\frac{\partial\varepsilon}{\partial n}-\varepsilon).
\end{equation}


The appearance of quadratic curvature terms suggests that we are dealing, actually, with the physical vacuum of the quantum field theory. This vacuum may produce particles --- quanta of the vacuum fluctuations of some fields. We assume that the classical fields that may also create particles, are absent.

Let us suppose that the universe was created from ``nothing''\cite{Vilenkin}. Most possible, it was initially empty, i.\,e., without particles. Then the question arises whether this vacuum persistent or it is just the initial state. Therefore we are in the situation when physical vacuum is able to create particles, but does not do this. It means that the creation law function is zero, but not all the coefficients ($\beta',\gamma'$) are zero,
\begin{equation} \label{PhiInv02}
	\Phi(\rm inv)=0, \quad |\beta'|+|\gamma'|\neq0,
\end{equation}
\begin{equation} \label{PhiInv02}
\frac{4}{3}\beta'R^0_0(2R^0_0-R)=\gamma'R^2.
\end{equation}
In the absence of particles $n=0$
\begin{equation} \label{G03}
	G^0[\rm part]=T^0[\rm part]=T[\rm part]=0.
\end{equation}
Let us have a look at the equation
\begin{equation} \label{dotlambda1}
\dot\lambda_1=-\frac{\varepsilon+p}{n}.
\end{equation}
It has quite a different solution, depending on the kind of matter particles. Namely, if $(\varepsilon+p)/n\to0$ for $n\to0$, then
\begin{equation} \label{lambda1const}
\lambda_1=const,
\end{equation}
but for the cosmic dust matter, $\varepsilon+p=\varphi(0)n$,
\begin{equation} \label{lambda1t0}
\lambda_1=-\varphi(0)(t-t_0).
\end{equation}
Consider first the non-dust matter. Then, for the general values $\beta',\gamma'\neq0$ one has 
\begin{equation} \label{lambda13}
\lambda_1=const, \quad R=\xi R^0_0, \quad \rightarrow \quad  (3\gamma'\xi^2+4\beta'(\xi-2))(R^0_0)^2=0. 
\end{equation}
\begin{eqnarray} \label{Tcr}
&&-6(\gamma-\gamma'\lambda_1)\dot R=0, \\ 
&&-12(\gamma-\gamma'\lambda_1)\frac{(\dot Ra^3)^{\dot{}}}{a^3}=0, \\ 
&&-(\gamma- \gamma'\lambda_1)\left(12\frac{\dot a}{a}\dot R 
+R(R-4R^0_0)\right)=0.
\end{eqnarray} 

In the case $\gamma\neq\gamma'\lambda_1$ ($\lambda_1$ is arbitrary), 
\begin{equation} \label{dotR0}
\dot R=0,
\end{equation}
and either we have the (flat) Milne universe ($R=R^0_0=0$) or $\xi=4$ --- de Sitter universe for $\beta+6\gamma'=0$.

Let $\gamma=\gamma'\lambda_1$, note that is not a special condition, but the solution for $\lambda_1$. Then, there exists the persistent vacuum with
\begin{equation} \label{C0}
\dot a^2+k=C_0 a^{\frac{4}{\xi-2}}, \quad C_0=const.
\end{equation}
We suspect that such a vacuum is unstable like the de Sitter vacuum in quadratic gravity, but do not study this problem here, mainly because the persistence of vacuum with respect to the creation of dust particles is not yet investigated. 

Let us now come to the dust matter case. First of all,
\begin{equation} \label{lambda1t02}
	\lambda_1=-\varphi_0(t-t_0), \quad \varphi_0=const.
\end{equation}
Then, the pregnancy implies (see Eq. (\ref{lambda13}))
\begin{equation} \label{lambda14}
R=\xi R^0_0, \quad \rightarrow \quad  (4\beta'(\xi-2)+3\gamma'\xi^2)(R^0_0)^2=0.
\end{equation}
The gravitational equations can be rewriting as follows (we omitted the tedious details),
\begin{eqnarray} \label{3final}
&&-6(\gamma-\varphi \gamma'(t-t_0))\dot  R=2\beta'\varphi(R-2R_0^0) +6\gamma'\varphi R, \\ 
&&12(\gamma-\varphi \gamma'(t-t_0))\frac{(\dot Ra^3)^{\dot{}}}{a^3}= 
4\beta'\varphi\frac{\dot a}{a}(R+2R_0^0) \\ 
&&+24\gamma'\varphi\dot R +36\gamma'\varphi\frac{\dot a}{a}R, \\ 
&&(\gamma-\varphi \gamma'(t-t_0))(12\frac{\dot a}{a}\dot R +R(R-4R_0^0)) \\ 
&&=
4\beta'\varphi\frac{\dot a}{a}(R-3R_0^0) +12\gamma'\varphi\frac{\dot a}{a}R.
\end{eqnarray}
Comparing the first and the third equation, we see that
\begin{equation} \label{RR}
R(R-4R_0^0)=0.
\end{equation}
Hence, either $R=0$ --- then $R^0_0=0$ and we have the (flat) Milne universe, or $R=4R^0_0$ --- then $\xi=4$, and it follows from the pregnancy condition (\ref{lambda14}) and the first equation that $\dot R=0$, i.\,e., we have the de Sitter universe

Note, that $\xi=4$ only if $\beta'+6\gamma'=0$. This means that for any other combination of $\beta'$ and $\gamma'$ the universe starts to produce dust particles immediately after its own birth.

\section{Discussion and conclusion}

In this paper we considered the application of Weyl gravitational theory to cosmology. By cosmology we understand the homogeneous and isotropic model with the Robertson-Walker metric. The Weyl gravity\cite{Weyl}, by definition, is invariant under the local conformal transformation. If one suppose the same type  of invariance in the case of Riemannian  geometry, then all the cosmological metrics (homogeneous and isotropic) would appear to be the vacuum solutions\cite{bde19,bdes20}.

We observed that the action integral for the matter fields is not obliged to be conformal invariant. It is its variation that has to obey such a requirement. In the Riemannian geometry, this conformal invariance condition leads to the tracelessness of energy-momentum tensor. However, in the Weyl geometry non-zero trace is allowed provided that there exists a direct interaction between the matter fields and the Weyl tensor, $A_\mu$, which we consider just as a part of the geometry. We found the corresponding self-consistency relation. 

We found that for the cosmological space-times strength tensor $F_{\mu\nu}=A_{\nu,\mu}-A_{\mu,\nu}=0$ (due to the high level of symmetry). Thus there exists the special gauge with $A_\mu=0$, i.\,e., the cosmological observers do not ``feel'' it at all, and one can safely think he is dealing with the Riemannian geometry.

For the matter fields we have chosen the perfect fluid and wrote the corresponding Lagrangian in the form suggested by J. R. Ray\cite{Ray}, where the particle number conservation law enters explicitly as a constraint. The Weyl gravitational Lagrangian contains the terms, quadratic in curvature. It is well known from the early 70s (of the last century) that the such terms describe the conformal anomaly responsible for the particle creation\cite{Parker69,GribMam70,ZeldPit71,HuFullPar73,FullParHu74,FullPar74}   
in one loop approximation of the quantum field theory. Therefore, we modified the corresponding constraint by allowing for the particle to be produced\cite{bde19,bdes20}. It is clear that this new term in the matter Lagrangian will produce some additional terms both in the Weyl current $G^\mu$ (the result of matter Lagrangian variation with respect to the Weyl vector $A_\mu$) and in the energy-momentum tensor $T^{\mu\nu}$. 

In order to determine the needed interaction between particles and the Weyl vector, we investigated the Lagrangian for single particle, moving in the given Weyl geometry (described by the metric tensor $g_{\mu\nu}$ and Weyl vector $A_{\mu\nu}$).

We found that the well known Lagrangian can be generalized by introducing the two functions of the new invariant $B=a_\mu u^\mu$, where $u^\mu$ is the four-velocity of particle. We derived the equations of motion and found that there are the conditions to be imposed in order to avoid contradiction. It appeared that in the case of cosmology they are satisfied automatically in our special gauge, making the latter quite physical. The insertion of these results into the perfect fluid Lagrangian was straightforward.

Since the whole theory was constructed as being conformal invariant it was reasonable to test the modified  constraint --- the rate of particle production
\begin{equation} \label{constr0}
	(nu^\mu)_{;\mu}\equiv\frac{(n\sqrt{-g}u^\mu)_{,\mu}}{\sqrt{-g}}=\Phi(\rm inv).
\end{equation}
The number density $n=\hat n/\Omega^3$, $u^\mu=\hat u^\mu/\Omega$, $\sqrt{-g}=\Omega^4\sqrt{-\hat g}$, where $\Omega$ is the conformal factor. Evidently $\Phi\sqrt{-g}$ must be conformal invariant. This result is of great importance, because it does not depend on the gravitational Lagrangian. In the absence of classical fields (what is already assumed by choosing the perfect fluid, consisting only of particles), the only choice for $\Phi(\rm inv)$ is the combination of exactly the same terms as in the Weyl gravitational Lagrangian, but, in general ``primed''  coefficients. Naturally, we restrict ourselves to the quadratic terms, in order not to go beyond the one-loop approximation of the quantum field theory.

Note, that in the Riemannian geometry (and, particularly, in General Relativity) one has no other choice for $\Phi(\rm inv)$, but the square of Weyl tensor. In 1977 Ya. B. Zel’dovich and A. A. Starobinskii\cite{ZeldStarob2} obtained the analogous result when considering the scalar particle production from the vacuum on the fixed background of the homogeneous and slightly anisotropic cosmological model. But now it becomes really fundamental.

Let us suppose that the universe was created from  ``nothing''\cite{Vilenkin} by some tunneling process, then, most probable, it emerged in a vacuum state. The question, therefore, arises, can  such a vacuum state persist or it is served just as an initial state. There are two quite different cases with quite different answers, the non-dust and dust ones, with the positive and zero hydrodynamic pressure, correspondingly. Of course, in both cases there exists the Milne Universe (locally flat space-time) as the solution, but it does not suit us because the emerging universe should be closed. Also there exists the de~Sitter solution for the special relation for the ``primed'' coefficients. In the general case such a vacuum exists in the non-dust case, but it does not exist in the dust case. This means, that the universe, just after emerging from the quantum vacuum foam, starts to produce dust particles. These dust particles do not interact with each other (by definition) and may represent the dark matter.

\ack

We are grateful to Yu. N. Eroshenko and A. L. Smirnov for stimulating discussions.

\appendix
\setcounter{section}{1}
\section*{Appendix A}\label{AA} 

Here we derive the part of the Weyl Current arising from the creation law in the matter integral, i.\, e., $G^\mu[\rm cr]$. By definition (see Eq. (\ref{TntGmucr}), 
\begin{equation} \label{TntGmucr2}
\int G^\mu[\rm cr](\delta A_\mu)\sqrt{-g}\,d^4x =\delta\!\!\int\!\lambda_1\Phi(\rm inv)\sqrt{-g}\,d^4x,
\end{equation}
where according to the creation law 
\begin{equation} \label{Phiinv}
\Phi(\rm inv)=\alpha_1'R_{\mu\nu\lambda\sigma}^{\mu\nu\lambda\sigma}
+\alpha_2'R_{\mu\nu}^{\mu\nu}
+\alpha_3'R^2+ \alpha_4'F_{\mu\nu}F^{\mu\nu}. 
\end{equation}
In order to obtain the left-hand-side of the vector gravitational equation, one should simply put $\lambda_1\equiv1$ and omit ``primes''.

Let us start with the $\alpha_1'$ term. One has
\begin{equation} \label{Ialpha11}
I[\alpha_1']=\int G^\mu[\alpha_1'](\delta A_\mu) =2\!\!\int\!\lambda_1 R_\mu^{\nu\lambda\sigma}(\delta R^\mu_{\nu\lambda\sigma}) \sqrt{-g}\,d^4x.
\end{equation}
The remarkable Palatini formula reads 
\begin{equation} \label{Ialpha1}
\delta R^\mu_{\nu\lambda\sigma}=\nabla_\lambda(\delta T^\mu_{\nu\sigma})  
-\nabla_\sigma(\delta T^\mu_{\nu\lambda})  
\end{equation}
Remembering that now $\delta \Gamma^\mu_{\nu\sigma}=\delta W^\mu_{\nu\sigma}$ and neglecting full derivatives, one gets
\begin{eqnarray} \label{Ialpha13}
I(\alpha_1')&=&-4\alpha_1'\int\left\{\nabla_\lambda(\lambda_1 R_\mu^{\phantom{\nu}\nu\lambda\sigma}) 
+2\lambda_1R_\mu^{\nu\lambda\sigma}A_\lambda\right\}(\delta W^\mu_{\nu\sigma}) \sqrt{-g}\,d^4x \nonumber \\
&=&-4\!\!\alpha_1'\int\! g_{\mu\mu'}g^{\nu\nu'}g^{\lambda\lambda'}g^{\sigma\sigma'} \nabla_\lambda(\lambda_1R^{\mu'}_{\phantom{\nu}\nu'\lambda'\sigma'})(\delta W^\mu_{\nu\sigma})  \sqrt{-g}\,d^4x. 
\end{eqnarray}
It follows, then
\begin{eqnarray} \label{Ialpha14} 
G^\mu[\alpha_1']&=&\alpha_1'g^{\lambda\lambda'}\left\{g^{\mu\sigma'}\nabla_\lambda(\lambda_1 R^{\nu'}_{\phantom{\nu}\nu'\lambda\sigma'})  +g^{\mu\sigma'}\nabla_\lambda(\lambda_1 R^{\nu'}_{\phantom{\nu}\sigma'\lambda'\nu'}) 
\right. \nonumber \\
&&-\left.g^{\nu'\sigma'}\nabla_\lambda(\lambda_1 R^{\mu}_{\phantom{\nu}\nu\lambda'\sigma'})\right\}. 
\end{eqnarray}
Since $R^{\nu'}_{\phantom{\nu}\nu'\lambda'\sigma'}=-2F_{\lambda'\sigma'}$ and    
$R^{\nu'}_{\phantom{\nu}\sigma'\lambda'\nu'}=-R_{\sigma'\lambda'}$,  
\begin{eqnarray} \label{Ialpha15} 
G^\mu[\alpha_1']&=&-2\alpha_1'g^{\lambda\lambda'} \left\{2g^{\mu\sigma'}\nabla_\lambda(\lambda_1F_{\lambda'\sigma'})  +g^{\mu\sigma'}\nabla_\lambda(\lambda_1 R_{\sigma'\lambda'}) 
\right. \nonumber \\
&&+\left.g^{\nu'\sigma'}\nabla_\lambda(\lambda_1 R^{\mu}_{\phantom{\nu}\nu'\lambda'\sigma'})\right\}. 
\end{eqnarray}
The last term can be transformed into
\begin{equation}\label{last}
g^{\mu\mu'}(\nabla_\lambda(\lambda_1 R_{\mu'\lambda'}+\nabla_\lambda(\lambda_1 F_{\mu'\lambda'})),
\end{equation}
and we get finally
\begin{equation}\label{final2}
G^\mu[\alpha_1']=-2\alpha_1'g^{\lambda\lambda'} g^{\mu\mu'} 
(2\nabla_\lambda(\lambda_1R_{\mu\lambda'}) -\nabla_\lambda(\lambda_1 F_{\mu'\lambda'})).
\end{equation}
In the case of cosmology, we are interested in, $F_{\mu\nu}\equiv0$, and, choosing our special gauge $A_\mu=0$, we have
\begin{equation}\label{final3}
G^\mu[\alpha_1']=-4\alpha_1'(\lambda_1R^{\mu\lambda})_{;\lambda}
\end{equation}
(the Weyl covariant derivatives becomes the metric ones).

It is a little bit simpler to calculate $G^\mu[\alpha_2']$, the result is
\begin{equation}\label{Galpha2}
G^\mu[\alpha_2']=-\alpha_2'(3g^{\mu\mu'}g^{\lambda\nu'} -g^{\mu\nu'}g^{\lambda\mu'} +g^{\mu\lambda}g^{\mu'\nu'})\nabla_\lambda(\lambda_1R_{\mu'\nu'}).
\end{equation}
For cosmology $R_{\mu'\nu'}=R_{\nu'\mu'}$, and in the special gauge $A_\mu=0$ one gets
\begin{equation}\label{Galpha2fin}
G^\mu[\alpha_2']=-\alpha_2'(2(\lambda_1R_{\mu\lambda})_{;\lambda} +(\lambda_1R)^{;\mu})
\end{equation}

The expression for $G^\mu[\alpha_3']$ is even more simple,
\begin{equation}\label{Galpha3}
G^\mu[\alpha_3']=-6\alpha_3'g^{\mu\lambda}g^{\kappa\sigma} \nabla_\lambda(\lambda_1R_{\kappa\sigma}),
\end{equation}
which in the case of cosmology becomes
\begin{equation}\label{Galpha3fin}
G^\mu[\alpha_3']=-6\alpha_3'(\lambda_1R)^{;\mu}.
\end{equation}

And, at last,
\begin{eqnarray} \label{Ialpha4}
I(\alpha_4')&=&2\alpha_4'\!\!\int\!F^\mu(\delta F_{\mu\nu})\sqrt{-g}\,d^4x \nonumber \\
&=&4\alpha_4'\!\!\int\!\left\{F^{\mu\nu}\sqrt{-g}(\delta A_\nu)_{,\mu}  - (F^{\mu\nu}\sqrt{-g})_{,\mu}\right\}(\delta A_\nu)\,d^4x. 
\end{eqnarray}
This gives for $G^\mu[\alpha_4']$
\begin{equation}\label{Galpha4fin}
G^\mu[\alpha_4']=4\alpha_4'(F^{\mu\nu})_{;\nu}.
\end{equation}
Note that the covariant derivative is the metric one. Evidently, this is zero in cosmology.

Combining all these, one gets, that for the special gauge  $A_\mu=0$ in cosmology,
\begin{eqnarray} \label{Gcrfin}
G^\mu[\rm cr]&=&-2(2\alpha_1' 
+\alpha_2')(\lambda_1R^{\mu\lambda})_{;\lambda} 
-(\alpha_2'+6\alpha_4')(\lambda_1R)^{;\mu}  \nonumber \\
&=&2\beta'((\lambda_1R^{\mu\lambda})_{;\lambda} -(\lambda_1R)^{;\mu})
-6\gamma'(\lambda_1R)^{;\mu}). 
\end{eqnarray}
It is easy to see, that $G^i[\rm cr]$ ($i=1,2,3$), and
\begin{equation}\label{G0crfin}
G^0[\rm cr]=2\beta'\dot\lambda_1(2R^0_0-R) -6\gamma'\dot\lambda_1R -6\gamma'\lambda_1\dot R.
\end{equation}
For gravitational equation ($\lambda_1\equiv1$), the term with $\beta'$ (appears in front of the Gauss-Bonnet term) disappears, and one is left simply with
\begin{equation}\label{G0crfinGB}
-6\gamma\dot R=G^0.
\end{equation}

\setcounter{section}{2}
\section*{Appendix B} \label{AB}

Here we will present a scheme for calculating the part of energy-momentum tensor $T^{\mu\nu[\rm cr]}$ that arises from the particle creation law. By definition,
\begin{equation}\label{Tcr}
\delta S_{\rm m}= 
-\frac{1}{2}\!\int\! T^{\mu\nu}[\rm cr](\delta g^{\mu\nu})\sqrt{-g}\,d^4x
=-\delta\!\int\!\lambda_1\Phi(\rm inv)\sqrt{-g}\,d^4x,
\end{equation}
\begin{equation} \label{PhiWeyl3}
\Phi(\rm inv) =\alpha_1'R_{\mu\nu\lambda\sigma}R^{\mu\nu\lambda\sigma}
	+\alpha_2'R_{\mu\nu}R^{\mu\nu}
	+\alpha_3'R^2+ \alpha_4'F_{\mu\nu}F^{\mu\nu}.
\end{equation}
Since we are interested only in cosmology, we will use the special gauge, $A_\mu=0$, from the very beginning, what is, surely, allowed now. Thus we are, actually, in the framework of the Riemannian geometry. Moreover, we will make use of the fact, that the Weyl tensor $C_{\mu\nu\lambda\sigma}=0$ for any Robertson-Walker metric.

We begin with the $\alpha'_1$--part. The routine procedure, with the help of  the Palatini formula\cite{Palatini} and algebraic symmetries of the curvature tensor, gives us
\begin{eqnarray} \label{Talpha1}
T^{\mu\nu}[\alpha_1']&=&-4\alpha_1'\left\{(\lambda_1 R^{\mu\kappa\nu\sigma})_{;\sigma;\kappa}
+(\lambda_1 R^{\nu\kappa\mu\sigma})_{;\sigma;\kappa}\right\}
\nonumber \\
&&+\alpha_1'\alpha_1\left\{R_{\kappa\sigma\lambda\delta} R^{\kappa\sigma\lambda\delta}g^{\mu\nu} 
- 4R_{\kappa\sigma\lambda}^{\phantom{abcd}\mu} R^{\kappa\sigma\lambda\nu}
 \right\}. 
\end{eqnarray}
The above expression can be considerably simplified. Le us star with the quadratic terms/ Using the evident relation  
\begin{equation} \label{rel}
R_{\mu\nu\lambda\sigma}R^{\mu\nu\lambda\sigma}=
C_{\mu\nu\lambda\sigma}C^{\mu\nu\lambda\sigma}
+2R_{\mu\nu}R^{\mu\nu} -\frac{1}{3}R^2
\end{equation}
and (much less evident) DeWitt's identity\cite{DeWitt}
\begin{equation} \label{DeW}
C_{\kappa\sigma\lambda}^{\phantom{abcd}\mu}C^{\kappa\sigma\lambda\nu}
=\frac{1}{2}C_{\kappa\sigma\lambda\delta}C^{\kappa\sigma\lambda\delta}
g^{\mu\nu}.
\end{equation}
What concerns the  linear part, it is not difficult to show that 
\begin{equation} \label{linear}
(\lambda_1R^{\mu\kappa\nu\sigma})_{;\sigma;\kappa}
=(\lambda_1R^{\nu\kappa\mu\sigma})_{;\sigma;\kappa}.
\end{equation}
Then, using the famous Bianchi identities and that $C_{\mu\nu\lambda\delta}=0$ for the cosmological metric, we arrive at the nice identity,
\begin{equation} \label{identity}
R^{\mu\kappa\nu\sigma}_{\phantom{abcde};\sigma}
=R^{\mu\nu;\kappa}-R^{\kappa\nu;\mu}
=\frac{1}{6}(R^{;\kappa}_{;\kappa}g^{\mu\nu} -R^{;\mu}g^{\nu\kappa}),
\end{equation}
which implies 
\begin{equation} \label{imply}
R^{\mu\kappa\nu\sigma}_{\phantom{abcde};\sigma;\kappa}
=\frac{1}{6}(R^{;\kappa}_{;\kappa}g^{\mu\nu} -R^{;\mu;\nu}).
\end{equation}
Finally, one gets for $T^{\mu\nu}[\alpha_1']$,
\begin{eqnarray} \label{Talpha1fin}
T^{\mu\nu}[\alpha_1']&=&-8\alpha_1'\left\{\frac{1}{2}\left({\lambda_1}^{;\kappa}_{;\kappa}(R^{\mu\nu} 
-\frac{1}{3}R g^{\mu\nu})
-{\lambda_1}^{;\nu}_{;\kappa}R^{\mu\kappa}
-{\lambda_1}^{;\mu}_{;\kappa}R^{\nu\kappa} \right. \right.\nonumber \\
&&+\left.\left.{\lambda_1}_{;\sigma;\kappa}R^{\kappa\sigma}g^{\mu\nu}
+\frac{1}{3}\lambda_1^{;\mu;\nu}R\right)
+\frac{1}{3}{\lambda_1}_{;\kappa}R^{;\kappa}g^{\mu\nu} \right.
\nonumber \\
&&-\left.\frac{1}{6}{\lambda_1}_{;\kappa}(R^{;\mu}g^{\nu\kappa}
+R^{;\nu}g^{\mu\kappa}) 
+\frac{1}{6}{\lambda_1}(R^{;\kappa}_{;\kappa}g^{\mu\nu} -R^{;\mu}_{;\nu})\right\}
\nonumber \\
&&+\alpha'_1\lambda_1\left\{-\frac{4}{3}RR^{\mu\nu}+\frac{1}{3}R^2g^{\mu\nu}\right\}.
\end{eqnarray}
In the same manner and using the same identities, one gets for $T^{\mu\nu}[\alpha_2']$,
\begin{eqnarray} \label{Talpha2fin}
T^{\mu\nu}[\alpha_2']&=&2\alpha_2'\left\{{\lambda_1}^{;\nu}_{;\kappa}R^{\mu\kappa} 
+{\lambda_1}^{;\mu}_{;\kappa}R^{\nu\kappa}
-{\lambda_1}^{;\kappa}_{;\kappa}R^{\mu\nu} 
-{\lambda_1}_{;\sigma;\kappa}R^{\sigma\kappa}g^{\mu\nu} \right.
\nonumber \\
&&+\frac{2}{3}({\lambda_1}^{;\nu}R^{;\mu} +{\lambda_1}^{;\mu}R^{;\nu} -2{\lambda_1}_{;\kappa}R^{;\kappa} g^{\mu\nu}) \nonumber \\ 
&&+\left.\frac{2}{3}\lambda_1(R^{;\mu;\nu} -R^{;\kappa}_{;\kappa}g^{\mu\nu})\right\}\!
+\alpha_1'\lambda_1\left\{\!-\frac{4}{3}RR^{\mu\nu}+\frac{1}{3}R^2g^{\mu\nu}\right\}\!. 
\end{eqnarray}
The most simple and straightforward is the calculation of $T^{\mu\nu}[\alpha_3']$, 
\begin{equation} \label{Talpha3fin}
T^{\mu\nu}[\alpha_3']=4\alpha_3'\left\{(\lambda_1R^{;\mu;\nu}) 
-(\lambda_1R)^{;\kappa}_{;\kappa}g^{\mu\nu}\right\}
+\alpha_3'\lambda_1\{R(Rg^{\mu\nu}-4R^{\mu\nu})\}. 
\end{equation}
Combining all these together and using coefficients ($\alpha',\beta\,\gamma'$), one gets finally
\begin{eqnarray} \label{Tcrfin}
T^{\mu\nu}[\rm cr]&=&4\beta'\left\{{\lambda_1}^{;\kappa}_{;\kappa}R^{\mu\nu} 
-{\lambda_1}^{;\nu}_{;\kappa}R^{\mu\kappa}
-{\lambda_1}^{;\mu}_{;\kappa}R^{\nu\kappa} 
+{\lambda_1}_{;\sigma;\kappa}R^{\sigma\kappa}g^{\mu\nu} \right.
\nonumber \\
&&\left.+\frac{2}{3}({\lambda_1}^{;\mu;\nu}R^{;\mu} -{\lambda_1}^{;\kappa}_{;\kappa}g^{\mu\nu})R\right\} \nonumber \\ 
&&+4\gamma'\left\{(\lambda_1R)^{;\mu;\nu} -(\lambda_1R)^{;\kappa}_{;\kappa}g^{\mu\nu} -\lambda_1R\!\left(\!R^{\mu\nu}-\frac{1}{4}Rg^{\mu\nu}\right)\!\right\}\!. 
\end{eqnarray}


\section*{References}
\begin{thebibliography}{99}

\bibitem{Vilenkin} Vilenkin A 1982 \emph{Phys. Lett. B}, {\bf 117} 25
\bibitem{Penrose} Penrose R 2014 \emph{Found. Phys.} {\bf 44} 873
\bibitem{Hooft} ’tHooft G 2015 \emph{arXiv} arxiv:1511.04427 [gr-qc]
\bibitem{Weyl} Weyl H 1918 \emph{Math. Zeit.} {\bf 2} 384
\bibitem{Ray} Ray J R 1972 \emph{J. Math. Phys.} {\bf 13} 1451
\bibitem{ZeldStarob} Zel’dovich Ya. B and Starobinskii A A 1972 \emph{Sov JETP} {\bf 34} 1159
\bibitem{ParkerFulling} Parker L and Fulling S A 1973 \emph{Phys. Rev. D} {\bf 7}  2357
\bibitem{GribMamMostep} Grib A A, Mamaev S G and Mostepanenko V M 1976 \emph{Gen. Relativ. Gravit.} {\bf 7} 535
\bibitem{Berezin} Berezin V A 1987 \emph{Int. J. Mod. Phys. A.} {\bf 2} 1591
\bibitem{bde19} Berezin V, Dokuchaev V and Eroshenko Yu 2019  \emph{ IJMPD} {\bf 28} 1941007  
\bibitem{bdes20} Berezin V, Dokuchaev V, Eroshenko Yu and Smirnov A  \emph{IJMPA} {\bf 35} 2040002
\bibitem{Ber14} Berezin V A 2014 \emph{Proc. 15th Russian Grav. Conf. ``RUSGRAV-15''} p. 58; arxiv:1404.3582
\bibitem{Parker69} Parker L 1969 \emph{Phys. Rev.} {\bf 183} 1057
\bibitem{GribMam70} Grib A A and Mamaev S G 1970 \emph{Sov. J. Nucl. Phys.}, {\bf 10} 722 
\bibitem{ZeldPit71} Zel'dovich Ya B and Pitaevsky L P 1971 \emph{Comm. Math. Phys.} {\bf 23} 185
\bibitem{HuFullPar73} Hu B L,  Fulling S A and Parker L 1973  \emph{Phys. Rev. D} {\bf 8} 2377
\bibitem{FullParHu74} Fulling S A, Parker L and Hu B L 1974 \emph{Phys. Rev. D} {\bf 10} 3905
\bibitem{FullPar74} Fulling S A and Parker L 1974 \emph{Ann. Phys. D} {\bf 87} 176
\bibitem{ZeldStarob2} Zel’dovich Ya. B and Starobinskii A A 1977 \emph{JETP Lett.} {\bf 26} 252
\bibitem{Palatini} Palatini A 1919 \emph{Rend. Circ. Mat. Palermo} {\bf 43}, 203
\bibitem{DeWitt} De Witt B S 1965 \emph{Dynamical theory of groups and fields} (Gordon and Bridge, New York) Chapters 13 and 16
 

\end{thebibliography}
\end{document}